# Enhanced broad band photoresponse of a partially suspended horizontal array of Silicon microlines fabricated on Silicon-On-Insulator wafers

Shaili Sett[1#*], Vishal Kumar Aggarwal[1#*], Achintya Singha[2*] and A. K. Raychaudhuri[1*]

[1]*Department of Condensed Matter Physics and Material Sciences,*
*S.N.Bose National Centre for Basic Sciences, JD Block, Sector –III, Kolkata 700106, India.*
[2]*Department of Physics,*
*Bose Institute, 93/1, Acharya Prafulla Chandra Road, Kolkata 700009, India.*
[#]Equal contribution from both authors
*E-mail: shaili.sett@bose.res.in, vka@bose.res.in, achintya@jcbose.ac.in and arup@bose.res.in

**Abstract:** We report a high Responsivity broad band photodetector working in the wavelength range 400nm to 1100nm in a horizontal array of Si microlines (line width ~1 μm) fabricated on a Silicon-on-Insulator (SOI) wafer. The array was made using a combination of plasma etching, wet etching and electron beam lithography. It forms a partially suspended (nearly free) Silicon microstructure on SOI. The array detector under full illumination of the device shows a peak Responsivity of 18A/W at 800nm, at a bias of 1V which is more than an order of magnitude of the Responsivity in a commercial Si detector (≤1A/W). In a broad band of 400nm to 1000nm the Responsivity of the detector is in excess of 10A/W. We found that the suspension of the microlines in the array is necessary to obtain such high Responsivity. The suspension isolates the microlines from the bulk of the wafer and inhibits carrier recombination by the underlying oxide layer leading to enhanced photoresponse. This has been validated through simulation. By using focused illumination of selected parts of a single microline of the array, we could isolate the contributions of the different parts of the microline to the photocurrent.

- **Introduction**

Research on high Responsivity optical detectors that have photo response over a broad wavelength region particularly in the infrared region is an important area of interest. Micro/nano structures that can be configured as Metal-Semiconductor-Metal (MSM) structure show promises for very high optical gain especially in low optical power range. Recently it has been established that single nanowire photodetectors made from semiconductors like Si [1], InGaAs [2] and Ge [3] shows ultrahigh sensitivity in a broad spectral range. In particular for Ge, a peak Responsivity of $10^7$A/W has been achieved. Self-powered single Ge nanowire (NW) photodetectors have been reported that work without an applied bias and show a Responsivity of $\sim 10^3$-$10^5$A/W in a broad spectral range [4, 5].While it is attractive to make single nanowire photodetectors with high Responsivity, they suffer from a bottleneck in the fabrication process that is tedious, with low output rate and is also incompatible with wafer level processing. A better alternative, albeit with compromise, is an array of such nanowire detectors that would be compatible with wafer level processing yet can give rise to high Responsivity [6], though not as high as a single NW detector. An example of such arrays with fairly high Responsivity reaching 1A/W in a narrow spectral range has been shown for metal oxide framework compound NWs [7]. The array of horizontal Si microline we report here is compatible with wafer level processing for fabrication yet they retain some advantages of single nanowire photodetector.

In the past few years SOI has emerged as an important material for novel device fabrication due to its wafer structure [8]. It is very valuable also for applications in microelectronics as the uppermost crystalline surface is used for electron transport. The buried layer of $SiO_2$ exists in between which acts as a dielectric layer that provides the insulating medium between device and wafer and also reduces the parasitic capacitance [8]. Recently it has been shown that a fully depleted SOI MOSFET photodetector with strong interface coupling shows a record Responsivity of $\sim 3\times10^4$ A/W in the visible region [9]. Si NWs with trapezoidal top, shift-line structures and Si island structures has been reported using electron beam lithography (EBL) and wet etching technique on SOI [10]. By the method of shifted mask patterns with combination of nanolithography and KOH anisotropic etching, Si nano-lines were fabricated of sub-10nm using SOI [11]. SOI substrates have also been used to fabricate photo-diodes and FET by using a combination of techniques like ion implantation, plasma etching and chemical etching through standard CMOS technology [12, 13]. A suspended SOI micro ring resonator has been fabricated using CMOS compatible technology for photothermal spectroscopy [14]. Si waveguides fabricated using standard lithography technique from SOI shows high confinement and non-linear optical effects [15].

In this work, we use SOI wafers to fabricate a broad band photodetector using a horizontal array of Si microlines. It is fabricated by a relatively simple process that uses a combination of wet etching, plasma etching and EBL. We have achieved a maximum Responsivity ~18A/W at 800 nm and Responsivity >10A/W over a broad spectral range of 400nm to 1000nm at a bias of 1V. Use of SOI for photonics application has a limitation that the supporting $SiO_2$ layer interface underneath the Si provides carrier recombination pathway that limits the optical response in such arrays. We show that by partially suspending the horizontal microlines (that constitutes the array) effectively mitigates the carrier recombination at the $Si/SiO_2$ interface.

- **Experimental Section**
  A. <u>Fabrication of device</u>

The SOI wafer has a top 2.5μm Si layer (P-doped with $\rho$~1-4Ω.cm, <100> oriented) with a 1μm buried $SiO_2$ layer supported on a 0.625 mm thick Si (B-doped with $\rho$~10Ω.cm). To achieve a Si building block of thickness ~ 100 nm, first we etch out the top layer of the fresh SOI wafer by Inductively Coupled Plasma Reactive Ion Etcher using a mixture of $Ar/CF_4$ gas [16]. After thinning down the wafer, we use EBL writing to draw the pattern of horizontal array of lines of specific dimensions (length~30μm and width ~1 μm) on double layer of PMMA resist, followed by lift-off after metallization using Au. This leads to a horizontal array of Au lines on top of the 100nm Si on the SOI wafer. The Au pattern acts as a mask for fabricating the Si microstructure using plasma etching. To protect the device, we use EBL to expose only the Au line array region of ~20μm length, while the rest of the device is coated in PMMA. We use anisotropic plasma etching to remove the exposed Si, which is not protected by the Au microlines. The Au on top of the Si nano-pattern is also etched out subsequently leaving an array of horizontal Si microlines of desired dimensions (~20μm in length) lying on top of $SiO_2$ sub-layer. The suspended microstructure is then achieved by selective etching using conc. HF solution, till the $SiO_2$ underneath the Si micro-blocks is etched out partially [16]. The $SiO_2$ should not be etched out completely as it provides mechanical support to the Si microlines. Finally we develop contact electrodes (Cr/Au~5/60nm) using EBL on the Si microline array. A schematic of entire process of device fabrication is shown in Figure 1.

  B. <u>Optical Measurements:</u>

The array of Si microlines was illuminated with a Xenon lamp (300-1100nm) coupled to a monochromator. The illumination area covers the whole array including the contact regions. The total light collection area of the array is ~$3.6 \times 10^{-6}$ $cm^2$. A schematic of the set-up is given in Figure 2. The illumination intensity of the light source was calibrated using a commercially available NIST calibrated Si photodetector. The illumination intensity per unit area was varied from 0.2 mW/cm$^2$ to 1.8 mW/cm$^2$.

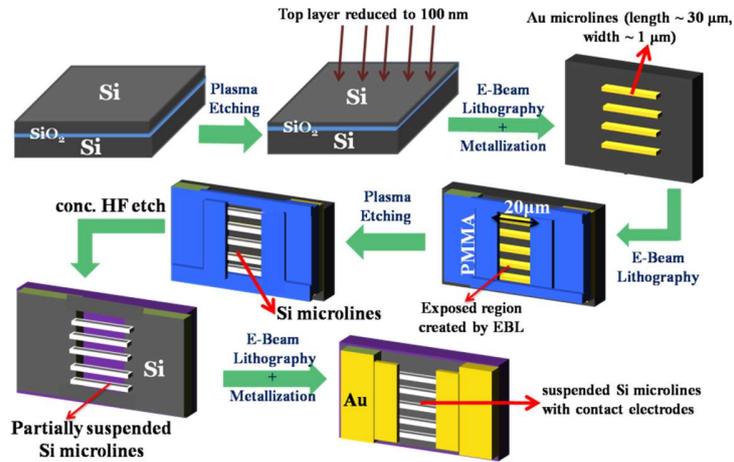

Figure 1. Schematic of device fabrication process.

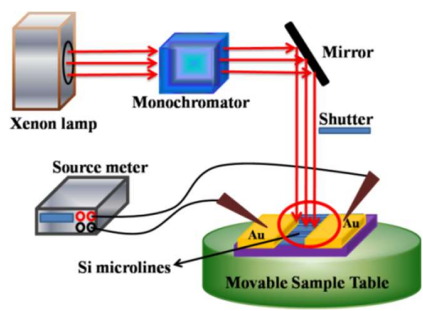

Figure 2. Schematic of the broad illumination of the Si microline array for photoresponse measurement.

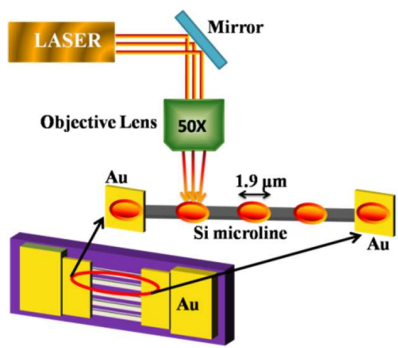

Figure 3. Schematic diagram showing the illumination of a single Si microline using a solid state laser through an objective lens. The light was focused on different parts of the microline as shown to measure the localized photo response.

Photoresponse of a single Si microline with focused illumination was performed with a solid state laser of wavelength 785nm. The focusing was done through a 50X objective lens. A schematic of the set-up is given in Figure 3. The FWHM of the laser is ~1.9μm. The length of the microline being 20μm, allows local illumination on different parts of the microline. The power of the laser was measured through a flux meter. In both cases, electrical measurements are taken through a sourcemeter with an automatic data acquisition using GPIB interface.

- **Results and Discussion**
  A. Characteristics of device

The Si microstructure device is shown in FESEM image of Figure 4(a). The inset of Figure 4(a) shows an optical microscope image of the complete array. In the device shown, there are 18 Si microlines that are connected in parallel to electrodes, each separated by a micrometer. The Si microlines have width ~ 1μm and length ~ 20μm (see Figure 4(a)) and are connected by Cr/Au electrodes. Figure 4(b) shows the magnified image of two microlines near one electrode. A partial suspension can be observed from the $60^0$ titled image. A schematic of a partially suspended Si microline due to partial etching of the SiO$_2$ layer is shown in Figure 4(c).

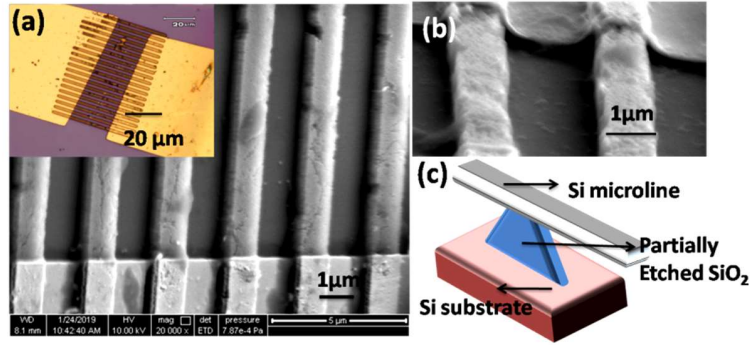

Figure 4. (a) FESEM image of array of Si microlines taken at a tilt of $60^0$. (Inset) Optical image of the entire array connected to Cr/Au electrodes. (b) Magnified image showing partially suspended Si microlines and (c) schematic of the partially suspended device created by etching of the underlying SiO$_2$ layer.

  B. Full array illumination

The current as a function of time at 1V bias, is measured when light is switched ON and OFF for different intensities in the range of 0.23mW/cm$^2$ to 1.8mW/cm$^2$ (at λ=650 nm). Typical *I-t* curves are shown in Figure 5(a). Photocurrent ($I_{ph}$) is defined as $I_{ph}(V) = I_{lig}\ (V) - I_{dark}(V)$, at a particular value of applied bias voltage (*V*). $I_{light}$ is the current from device when illumination is ON and $I_{dark}$ is the current from device when the illumination is OFF at the same bias. In addition to bias, $I_{ph}$ also depends on wavelength and intensity of the illumination.

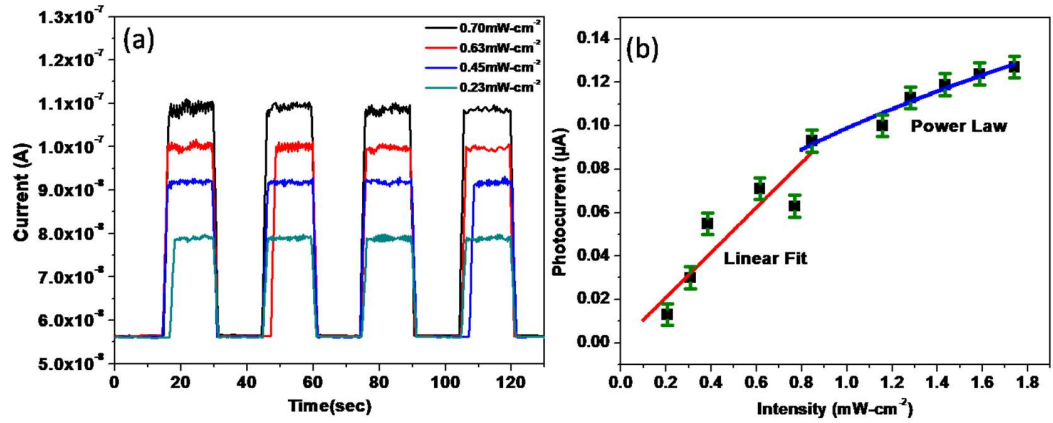

Figure 5. (a) Current as a function of time at different intensities. (b) Photocurrent as a function of Intensity at bias voltage 1V.

Photocurrent as a function of Intensity is studied at a bias of 1V at wavelength of 900nm. As illumination power $P$ increases $I_{ph}$ increases because of enhancement in electron-hole pair generation. The behavior of photocurrent ($I_{ph}$) with Intensity ($\mathcal{J}$) is shown in Figure 5(b). The Power ($P$) of incident light falling on the array has been calculated from the illumination intensity ($\mathcal{J}$) using the expression, $P = (\mathcal{J} \times A)$, where $A$ is the area of the array over which incident light is collected, given by $A = (n \times l \times b)$, where $n$ is number of Si microlines in array, $l$ is length of Si microlines between two electrodes and $b$ is the width of a Si microline. Initially at low intensity of upto 0.8mW/cm$^2$, photocurrent increases linearly with intensity. At higher intensities, the photocurrent has sub-linear dependence on intensity, indicating presence of trap states near the conduction band tails. The higher intensity data is fitted using power law, $I_{ph} \propto \mathcal{J}^\alpha$ where exponent $\alpha$ was found to be ~0.47. The exponent $\alpha$ depends on the factors like, process of generation of electron-hole pairs and trapping and recombination of charge carriers in the device [17]. If the distribution of trap states near the Fermi level depends exponentially on the density of states at the conduction band edge, then the characteristic energy scale $E^*$ associated with the trap states is related to the power law exponent as [1,3,7],

$$\alpha = \frac{E^*}{(E^* + k_B T)} \qquad (1)$$

This gives an energy scale of $E^* = 22$meV. The value of $E^*$ suggests that these states are within the donor levels ($\sim 45$meV) formed by P-dopant in Si below the conduction band. The process of recombination of charge carriers which results in photocurrent is effectively controlled by the localized trap states at illumination intensity higher than 0.8mW/cm$^2$.

One relevant parameter that characterizes photodetectors is Responsivity, $R = \left(\frac{I_{ph}}{P}\right)$. We measured $R$ for different values of bias $V$ over a wavelength range of 400nm to 1100nm as shown in Figure 6(a). The peak $R$ of the Si microline array increases from $\sim$5A/W for $V$=50mV to $\sim$18A/W at $V$=1V as shown in Figure 6(b). At the maximum bias (1V) in which we have operated the device, $R$>10A/W for band width of 400nm-1000nm and reaches a maximum of 18A/W at 800nm. This is the wavelength range where conventional Si photodetectors also reach a maximum. The Responsivity of the Si microline detector (photoconductor) is $\sim$30 times larger than commercially available Si detectors (for peak wavelength), which are a p-n junction type photodetectors. (Note: We have assumed 100% absorption of the incident light. This gives a lower estimate of the Responsivity as the reflectivity $\leq$ 40% is not accounted for. This will reduce the actual power absorbed $P$ by a factor of nearly 2.5 and thus making the effective Responsivity higher by the same factor.)

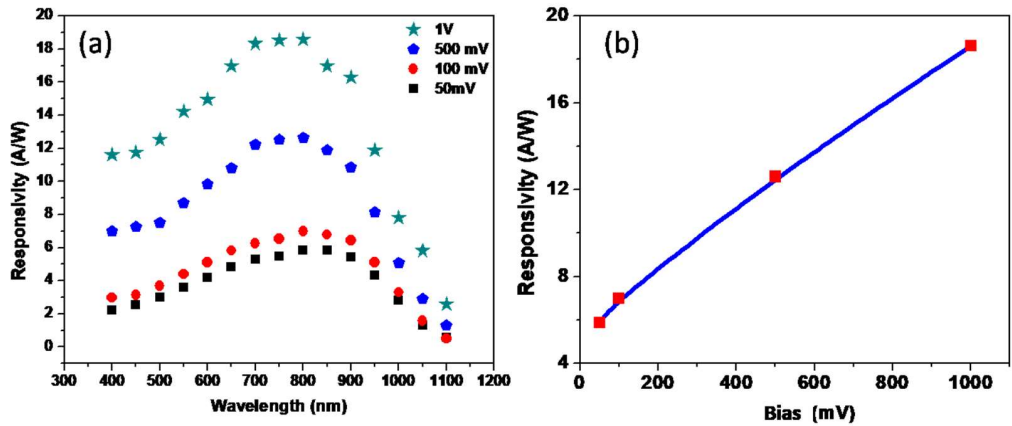

Figure 6. (a) Responsivity as a function of wavelength at different values of applied bias. (b) Peak Responsivity (λ=800nm) as function of applied bias.

*C*. Illumination of a single Si microline

In addition to measuring the photoresponse in the whole array of microlines with a broad illumination, we also investigated photoresponse of a single microline with a focused illumination. The motivation to study a single microline is to investigate the variability of photoresponse from different regions of a given microline that is a constituent of the array. As shown in Figure 3, the microline was illuminated by a focusing the light from a 785 nm solid state laser. We have focused the laser at the two contacts and at three equally spaced spots in the Si microline (as shown in Figure 7) where we have shown $I_{ph}$ (measured with a bias of 1V) as a function of position of the laser spot for different intensities (see Table 1 for details). The uncertainty in current response (shown as error bar) comes from the current noise in the detector. We observe that while photocurrent is generated from different regions of the microline, the generation efficiency differs in each portion. The observed $I_{ph}$ when the contact region is illuminated is higher than that when the centre of the microline is illuminated. Generation of photocurrent from the contact is characteristic of a Schottky Photocurrent (SPC) behavior where the width of depletion region is reduced as well as the Schottky barrier height (SBH) is lowered due to excess carrier density created by illumination [1,3,18]. The $I_{ph}$ from contact illumination at the maximum intensity ($1.74 \times 10^9$ W-m$^{-2}$) reaches ≈125nA which is larger than $I_{ph}$ ≈75nA when the central region alone is illuminated. The lowering of SBH ($\Delta\phi$) due to creation of carriers by illumination at the contact region can be estimated from the change in the chemical potential given by [3, 7],

$$|\Delta\phi| = \frac{k_B T}{q} \ln\left(\frac{n_0 + \Delta n}{n_0}\right) \tag{2}$$

where, $n_0$ is the carrier concentration in dark and $\Delta n$ is the additional photocarrier generation. Here we assume that all carriers that are photogenerated get collected, which sets an upper limit on $\Delta\phi$. From conductivity change upon illumination, we can calculate the fractional change in carrier density $\frac{\Delta n}{n_0}$. Using Eqn. 1, we get $\Delta\phi$ ≈10meV. The SBH in a Si nanowire with Cr/Au contact is known to be $\phi$ ≈75meV [1]. Assuming that it is similar in the case of a Si microline, the observed $I_{ph} \approx$ 125nA would correspond to a fractional decrease in the barrier height, $\frac{\Delta\phi}{\phi_0}$ ≈13% upon illumination.

Table 1. Photocurrent under illumination (at wavelength 785nm and Intensity $1.7 \times 10^9$ W-m$^{-2}$) at different regions of the Si microline.

| Position of laser spot | $I_{ph}$ (nA) |
|---|---|
| Contact 1 | 104±14 |
| A (between contact 1 and centre) | 52±14 |
| B (centre) | 76±14 |
| C (between centre and contact 2) | 49±14 |
| Contact 2 | 146±14 |

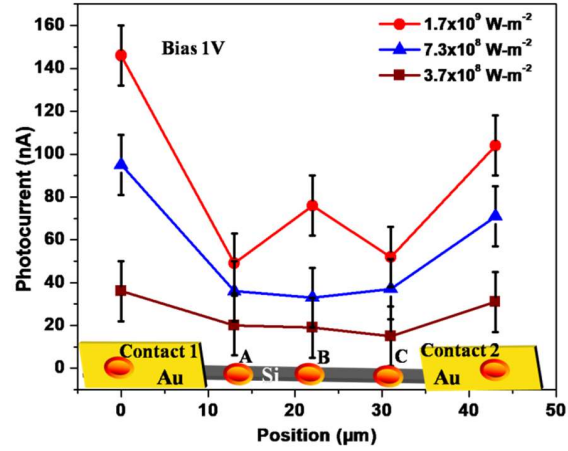

Figure 7. Photocurrent as a function of position of laser spot on a single microline as shown by schematic below.

The ratio of photocurrent generated from centre of the microline $(I_{Ph})_{cen}$ to the contacts $(I_{Ph})_{contact}$ is given in Table 2 for different intensities. As intensity increases, this ratio also increases. At highest intensity, this ratio ~0.6, which indicates that the photocurrent generated from the microline itself is quite high. Even though the centre of the microline is ~ 9μm away from the metal-semiconductor (MS) contact we get substantial photocurrent from a single microline. (Note: the typical recombination length of the photogenerated carriers in Si NW are~ 3-4μm [19] and the depletion width is ≤ 1μm at the contacts [20]).

One of the factors that give rise to relatively large $I_{ph}$ when the middle of the microline is illuminated is the partial suspension of the microlines by isolating them from the underlying oxide layer. We find that when the microlines are not suspended, the response is very low. We propose that partial suspension of the microlines inhibits recombination of carriers by the underlying oxide layer. This enhances the recombination lifetime of the carriers leading to enhanced response. We carry out a simulation to support our proposal. The simulation establishes the efficacy of the suspension in enhancing the photoresponse.

Table 2. The photocurrent generated locally from a single Si microline

| Intensity (x10$^8$W-m$^{-2}$) | $(I_{Ph})_{cen}$ (nA) | $(I_{Ph})_{contact}$ (nA) | $r \equiv \dfrac{(I_{Ph})_{cen}}{(I_{Ph})_{contact}}$ |
|---|---|---|---|
| 17.4 | 76±14 | 125±35 | 0.6±0.19 |
| 7.38 | 33±14 | 83±16 | 0.4±0.15 |
| 3.74 | 19±14 | 67±27 | 0.28±0.21 |

C. Simulation of Device

The enhanced performance of the microline array photodetector was achieved when the recombination pathways in the oxide layer of the SOI wafer was reduced by partially suspending the microline. To validate this proposal, we performed simulation of the photocurrent generation in a microline using COMSOL Multiphysics software. The device geometry used is a supported and a partially suspended Si microline as shown in Figure 8. We have varied the width of the oxide layer ($W$), keeping the thickness fixed (200nm). For the simulation we have used an incoming electromagnetic (EM) wave of wavelength 785nm, with 1W power incident perpendicularly on the photodetector.

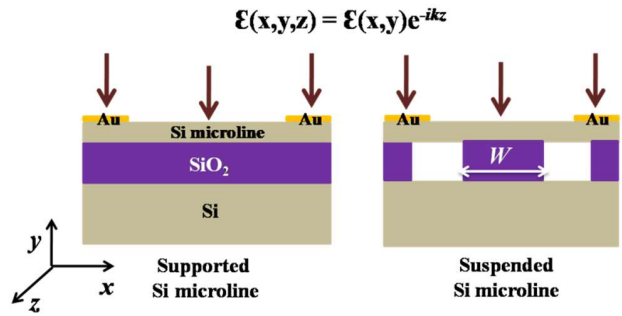

Figure 8. Device geometry of supported and partially suspended Si microline used for simulation.

To solve for the device current, we use Equations 3 and 4 that define the electron ($J_n(r,t)$) and hole ($J_p(r,t)$) current densities used in the simulation.

$$J_n(r,t) = qn\mu_n \nabla \varphi_c + \mu_n k_B T K\left(\frac{n}{N_c}\right)\nabla n + \left(\frac{nq}{T}\right)D_{n,th}\nabla T \qquad (3)$$

$$J_p(r,t) = qp\mu_p \nabla \varphi_v + \mu_p k_B T K\left(\frac{p}{N_v}\right)\nabla p - \left(\frac{pq}{T}\right)D_{p,th}\nabla T \qquad (4)$$

where, $n, p$ are the electron and hole carrier concentrations respectively, $\mu_n, \mu_p$ are the electron and hole mobilities respectively, $\nabla\varphi_c, \nabla\varphi_v$ are the gradient of potential terms of electron and hole respectively, $k_B$ is the Boltzmann constant, $T$ is the temperature, $N_c, N_v$ are the effective density of states in conduction band and valence band respectively and $D_{n,th}, D_{p,th}$ are thermal diffusion coefficients. $K$ is related to the Fermi-Dirac Integral as, $K(\alpha) = \dfrac{\alpha}{F_{-\frac{1}{2}}\left(F_{\frac{1}{2}}^{-1}(\alpha)\right)}$ where, $F_j(\alpha)$ is the $j^{th}$ order Fermi-Dirac integral and $\alpha$ is the dummy index which can be $\left(\dfrac{n}{N_c}\right)$ or $\left(\dfrac{p}{N_v}\right)$. To solve the model we incorporate Poisson's equation and current continuity equations given by,

$$\frac{\partial n}{\partial t} = \frac{1}{q}(\nabla \cdot J_n) - U_n \qquad (5)$$

$$\frac{\partial p}{\partial t} = \frac{-1}{q}(\nabla \cdot J_p) - U_p \qquad (6)$$

where, $U_n = \Sigma R_{n,i} - \Sigma G_{n,i}$ is the net electron recombination rate from all the electron generation ($G_{n,i}$) and recombination mechanisms ($R_{n,i}$). Similarly $U_p$ is the net hole recombination rate from all the generation $G_{p,i}$ and recombination mechanism $R_{p,i}$. The recombination of carriers occurs through Shockley-Read-Hall mechanism. After setting up the semiconductor interface, we use EM wave interface to simulate the photocurrent in the device. We then apply the indirect optical transition mechanism (that uses the refractive index of Si at 785 nm, *n* and *k*) to simulate the optical absorption in the Si microline. The photo-generation due to absorbed photons is then converted to the carrier generation rate $G_{iot}$ defined as,

$$G_{iot} = \left(\frac{1}{\hbar}\right)\epsilon_o n k \mathcal{E}^2 \qquad (7)$$

where, $\epsilon_o$ is free space permittivity and $\mathcal{E}$ is the normal electric field applied.

As we decrease the width of the oxide layer *W* and increase the extent of device suspension, there is a systematic increase in the photocurrent as shown in Table 3. The important quantity that causes this change is the net carrier recombination rate $R_n$ which decreases as we increase the suspension of Si microline. From the surface plot of Figure 9, (see color bar) we observe that recombination rate per unit volume peaks at the region where Si is supported by the oxide layer which decreases on reduction of the underlying oxide layer. The recombination rate in the Si microline as a function of the length of the oxide underneath the micrline is shown in Figure 10. The relative enhancement in the photocurrent on reduction of oxide length *W* is also shown in Figure 10. The changes in *W* has been shown as relative ratio with respect to the Si microline length. There is ~60% increase of photocurrent as the underneath oxide is reduced from 100% (fully supported) to 20% (almost fully suspended). There is a saturation of the recombination rate at higher oxide support. The graph establishes that as we increase the degree of suspension, recombination reduces which results in enhancement of photocurrent.

Figure 11 shows the Electric field distributions in two devices upon illumination (as obtained in the simulation) at a bias of 5V. The direction and magnitude of the electric field are shown as arrows in the image. The current density is shown in Figure 12. There is a large increase in the device current density as the supported device is made into a suspended device. The direction of current flow is shown as arrow plot. There is very low current density in the oxide layer and underlying Si in case of the suspended structure in comparison to the supported one. The simulation clearly establishes the efficacy of the suspension in enhancing the photoresponse and the gradual enhancement in the photocurrent as the degree of suspension is increased.

Table 3. Increment in Photocurrent with change in width of oxide layer beneath Si microline.

| $W$ (μm) | $I_{dark}$ (nA) | $I_{ph}$ (nA) |
|---|---|---|
| 0.2 | 0.44 | 144.8 |
| 1 | 0.45 | 138.7 |
| 2 | 0.46 | 131.7 |
| 3 | 0.48 | 125.4 |
| 4 | 0.50 | 119.6 |
| 5 | 0.54 | 112.8 |

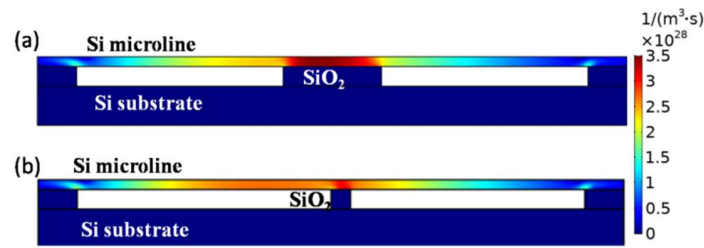

Figure 9. Surface plot of the carrier Recombination rate per unit volume in (a) partially supported Si microline and (b) in a highly suspended Si microline in the presence of illumination.

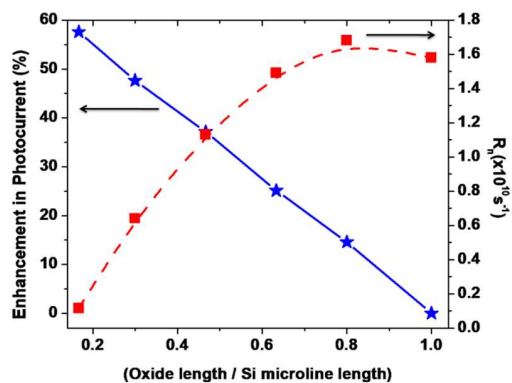

Figure 10. Dependence of the enhancement of photocurrent (over dark current) and the effective carrier recombination rates on the length of the underlying $SiO_2$ layer that supports the microlines. The length of the oxide layer has been plotted as a ratio of the Si microline length.

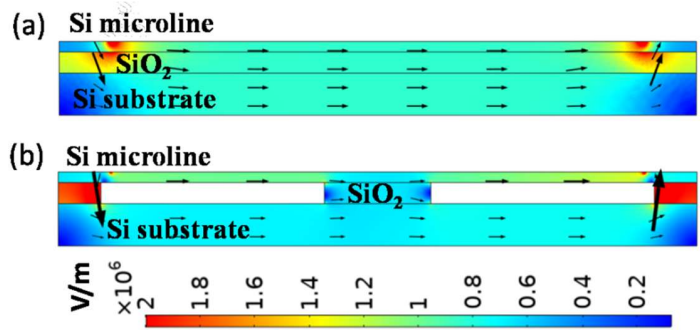

Figure 11. Surface plot of Electric Field in (a) a supported microline and (b) a suspended microline. The arrows show the directions of the Electric field.

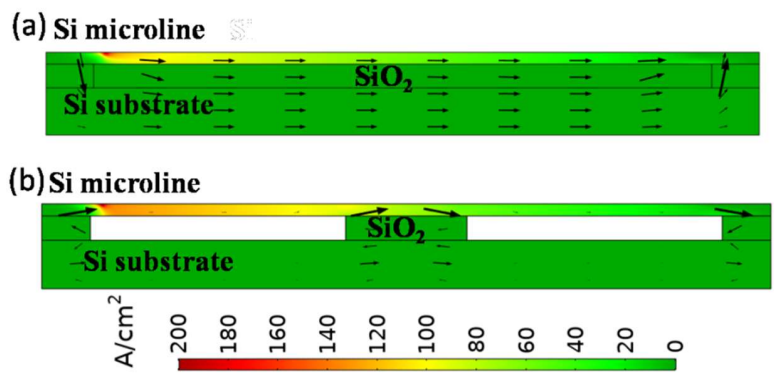

Figure 12. Current densities in (a) a supported microline and (b) a suspended microline. The arrows show the directions of current.

- **Conclusions**

In summary, we have developed a high responsivity Si microline array photodetector from SOI wafer using a top-down method that is compatible with wafer scale processing. It requires a combination of dry plasma etching, wet etching and EBL to form a uniform horizontal array of well-defined Si microlines that are effectively isolated from the oxide layer below through partial suspension. The Responsivity is at least an order higher than commercially available bulk Si detectors in the same spectral range. The structure itself provides the necessary conditions for maximizing radiation collection area and reducing recombination pathways. Using simulation we establish the efficacy of enhancing the photoresponse on suspending the microline that removes recombination on the underlying oxide layer. Focused illuminations of a single Si microline at selected spots show that substantial photocurrent is generated from a single Si microline although there is Schottky photo-diode behavior at the contacts.

**Author Information**

Corresponding Author

*E-mail: arup@bose.res.in

Notes:

The authors declare no competing financial interest.

**Acknowledgement**

The authors acknowledge financial support from Department of Science and Technology, Govt. of India for the sponsored project Theme Unit for Nanodevice Technology (Grant SR/NM/NS09/2011(G) funded by the Nanomission). AKR acknowledges financial support from Science and Engineering Research Board, Government of India as a J.C. Bose Fellowship (SR/S2/JCB-17/2006). The authors also thank Prof. Barnali Ghosh for use of the photoconduction set-up.

**References**

[1] Das K, Mukherjee S, Manna S, Ray S K and Raychaudhuri A K 2014 *Nanoscale.* **6** 11232-11239

[2] Tan H, Fan C, Mal L, Zhang X, Fan P, Yang Y, Hu W, Zhou H, Zhang X, Zhu X and Pan A 2016 *Nano-Micro Lett.* **8** 29-35

[3] Sett S, Ghatak A, Sharma D, Kumar G V Pavan and Raychaudhuri A K 2018 *J. Phys. Chem. C.* **122** 8564-8572

[4] Sett S, Sengupta S and Raychaudhuri A K 2018 *Nanotech.* **29** 445202.

[5] Mukherjee S, Das K, Das S and Ray S K 2018 *ACS Photonics.* **5** 4170

[6] Tran D P, Macdonald T J, Wolfrum B, Stockmann R, Nann T, Offenhausser A and Thierry B 2014 *Appl. Phys. Lett.* **105** 231116

[7] Basori R, Das K, Kumar P and Raychaudhuri A K 2013 *Appl. Phys. Lett.* **102** 061111-1-061111-4

[8] Ploeul A and Kraeuter G 2000 *Solid-State Electronics* **44** 775-782

[9] Deng J, Shao J, Lu B, Chen Y and Zaslavsky A 2018 *IEEE J. Elec. Devices Soc.* **6** 557-564

[10] Han W, Yang X, Wang Y, Yang F and Yu J 2008 *IEEE* 146-148

[11] Kurihara K, Namatsu H, Nagase M and Makino T 1996 *Jpn. J. Appl. Phys.* **35** 6668-6672

[12] Geis M W et al. 2007 *Optics Express* **15** 16886

[13] Khamaisi B, Vaknin O, Shaya O and Ashkenasy N 2010 *ACS Nano* **4** 4601-4608

[14] Vasiliev A, Malik A, Muneeb M, Kuyken B, Baets R and Roelkens G 2016 *ACS Sens.* **1** 1301-1307

[15] Sederberg S and Elezzabi A Y 2014 *ACS Photonics* **1** 576-581

[16] Pescini L, Tilke A, Blick R H, Lorenz H, Kotthaus J P, Eberhardt W and Kern D 1999 *Nanotech.* **10** 418-420

[17] Rose A 1978 Concepts in Photoconductivity and Allied Problem (Krieger, New York)

[18] Ahn Y, Dunning J and Park J 2005 Nano *Lett*. **5** 1367-1370

[19] Kato S, Kurokawa Y, Miyajima S, Watanabe Y, Yamada A, Ohta Y, Niwa Y and Hirota M 2013 *Nanoscale Research Lett*. **8** 361

[20] Sze S M 1981 *Physics of Semiconductor Devices* 2nd ed. (Wiley, London)